\documentclass[useAMS,usenatbib]{mn2e}
\usepackage{graphicx}
\usepackage{color}
\newcommand{\etal}{et al.~}
\newcommand{\rhocrit}{\rho_c}
\newcommand{\rhorms}{\rho_{\rm rms}}
\newcommand{\xoff}{x_{\rm off}}
\newcommand{\cvir}{c_{\rm vir}}
\newcommand{\Rvir}{R_{\rm vir}}
\newcommand{\Mvir}{M_{\rm vir}}
\newcommand{\rs}{r_{\rm s}}
\newcommand{\Jvir}{J_{\rm vir}}
\newcommand{\Vvir}{V_{\rm vir}}
\newcommand{\Nvir}{N_{\rm vir}}
\newcommand{\Ms}{M_{\rm s}}
\newcommand{\cM}{$\rm c_{\rm vir}-M_{\rm vir}$~}

\newcommand{\mnras}{MNRAS}
\newcommand{\apj}{ApJ}
\newcommand{\apjl}{ApJ}
\newcommand{\apjs}{ApJS}

\def \LCDM {\ifmmode \Lambda{\rm CDM} \else $\Lambda{\rm CDM}$ \fi}
\def \kms {\ifmmode  \,\rm km\,s^{-1} \else $\,\rm km\,s^{-1}  $ \fi }
\def \kpc {\ifmmode  {\rm kpc}  \else ${\rm  kpc}$ \fi  }  
\def \Msun {\ifmmode M_{\odot} \else $M_{\odot}$ \fi} 
\def \hMsun {\ifmmode h^{-1}\,\rm M_{\odot} \else $h^{-1}\,\rm M_{\odot}$ \fi}

\title[Mass and redshift dependence of concentration, spin and shape]
{The redshift evolution of \LCDM halo parameters: concentration, spin and shape}
\author[J. C. Mu\~noz-Cuartas \etal]
{J.C. Mu\~noz-Cuartas$^{1}$\thanks{Email : jcmunoz@aip.de}, A.V. Macci{\`o}$^{2}$, S.Gottl\"{o}ber$^{1}$,
A.A. Dutton$^{3}$\thanks{CITA National Fellow}\\
$^{1}$ Astrophysikalisches Institut Potsdam, An der Sternwarte 16, 14482
Potsdam, Germany. \\
$^{2}$ Max-Planck-Institut F\"{u}r Astronomie, K\"{o}ningstuhl 17, 69117
Heidelberg, Germany.\\
$^{3}$ Department of Physics and Astronomy, University of Victoria, Victoria, B.C., V8P 5C2, Canada.}
\begin{document}

\date{Accepted 2010 September 13. Received 2010 September 13; in original form
  2010 July 1}

\pagerange{\pageref{firstpage}--\pageref{lastpage}} \pubyear{2010}

\maketitle

\label{firstpage}

\begin{abstract}
  We present a detailed study of the redshift evolution of dark matter
  halo structural parameters in a $\Lambda$CDM cosmology. We study the
  mass and redshift dependence of the concentration, shape and spin
  parameter in N-body simulations spanning masses from $10^{10}$
  \hMsun\ to $10^{15}$ \hMsun\ and redshifts from 0 to 2. We present a
  series of fitting formulae that accurately describe the time
  evolution of the concentration--mass ($\rm c_{\rm vir}-M_{\rm vir}$)
  relation since $z=2$. Using arguments based on the spherical
  collapse model we study the behaviour of the scale length of the
  density profile during the assembly history of haloes, obtaining
  physical insights on the origin of the observed time evolution of
  the \cM relation. We also investigate the evolution with redshift of
  dark matter halo shape and its dependence on mass. Within the
  studied redshift range the relation between halo shape and mass can
  be well fitted by a redshift dependent power law. Finally we
  show that although for $z=0$ the spin parameter is practically mass
  independent, at increasing redshift it shows a increasing
  correlation with mass.
\end{abstract}



\section{Introduction}

Observational evidence (e.g. Spergel \etal 2007; Komatsu \etal 2009)
favours the hierarchical growth of structures in a universe dominated
by cold dark matter (CDM) and dark energy ($\Lambda$), the so called
\LCDM universe.  Within this paradigm dark matter collapses first into
small haloes which accrete matter and merge to form progressively
larger haloes over time.  Galaxies are thought to form out of gas
which cools and collapses to the centres of these dark matter haloes
(e.g.  White \& Rees 1978).  In this picture the properties of
galaxies are expected to be strongly related to the properties of the
dark matter haloes in which they are embedded (e.g.  Mo, Mao \& White
1998, Dutton \etal 2007).

It has been shown by several studies that the structural properties of
dark matter (DM) haloes are dependent on halo mass: for example higher
mass haloes are less concentrated (Navarro, Frank \& White 1997,
hereafter NFW; Eke \etal 2001; Bullock \etal 2001a; Kuhlen \etal 2005;
Macci\`o \etal 2007; Neto \etal 2007; Gao \etal 2008; Duffy \etal
2008; Macci\`o, Dutton, \& van den Bosch 2008 (hereafter M08); Klypin
\etal 2010), and are more prolate (Jing \& Suto 2002; Allgood \etal
2006; Gottl\"ober \& Yepes 2007; Bett \etal 2007; Macci\`o \etal 2007;
M08) on average. The situation is less clear for the spin parameter,
at $z=0$ there seems to be no mass dependence (Macci\`o \etal 2007;
M08) or at least a very weak one (Bett \etal 2007), while for
increasing values of the redshift a possible mild correlation between
spin and mass seems to arise (Knebe \& Power 2008).

In M08 the properties of DM haloes were studied in \LCDM universes
whose parameters were fixed by the one, three and five-year release of
the WMAP mission (WMAP5; Komatsu \etal 2009).  In that study the
attention has been focused on the structural parameters of virialized
haloes and their correlations at the present epoch, $z=0$.  In this
work we extend this previous analysis to higher redshifts and study
how the scaling relations of DM haloes change with time.

As in M08 we use a large suite of N-body simulations in a WMAP5
cosmology with different box sizes to cover the entire halo mass range
relevant for galaxy formation: from $10^{10} \hMsun$ (haloes that host
dwarf galaxies) to $10^{15} \hMsun $ (massive clusters).  We use these
simulations to investigate the evolution of concentrations, spin
parameter and shapes of dark matter haloes through cosmic time.
 
Similar studies have been already conducted in the past, mainly using
lower numerical resolution and/or a smaller mass range (but with few
recent exceptions).

Navarro, Frank \& White (1997) proposed that the characteristic
density of dark matter haloes was directly proportional to the density
of the universe at time of formation, making possible to connect today
properties of the dark matter density profile to the halo formation
history and to the evolution of the expanding universe.  This idea was
then expanded by Wechsler \etal (2002), who found a clear connection
between the mass growth of dark matter haloes and the definition of
the formation time; connecting directly the growth history of dark
matter haloes to the evolution of their concentration parameter.

In a series of papers, Zhao \etal (2003a, 2003b, 2009) have
re-addressed the problem of the evolution of dark matter halo density
profile and the mass accretion history. Zhao's main result was the
finding of a correlation between $\rs$ and the characteristic mass of
dark matter haloes, $\Ms$, defined as the mass inside $\rs$. Thanks to
this correlation they were able to model the time evolution of the
concentration parameter in a cosmology-free fashion.

A comparison of the different approaches to predict dark matter halo
concentrations was performed by Neto \etal (2007).  They made a
detailed comparison of the Wechsler \etal (2002) and Zhao \etal (2003)
models for the time evolution of dark matter halo masses and their
resulting predictions for halo concentration.  Neto \etal (2007) found
that although these models could match the average concentration
reasonably well, they performed very poorly in many cases, because
their models for halo mass evolution were not able to satisfactorily
reproduce "real" mass growth histories from N-body simulations.

However, the evolution of dark matter halo properties does not reduce
to the concentration parameter only; halo shape and spin parameter are
also important quantities that could influence the properties of the
hosted galaxy. Allgood \etal (2006) studied the mass, radius, redshift
and cosmology dependence (via variations of $\sigma_8$) of the shape
of dark matter haloes; while the environment dependence of the shape
has been addressed in Hahn \etal (2007a,b).  The distribution of the
spin parameter of dark matter haloes has been studied in several works
(e.g. Bullock \etal 2001b; Macci\`o \etal 2007; Bett \etal 2007; M08;
Knebe \& Power 2008; Davis \& Natarajan 2009; Antonuccio-Delogu \etal
2010) as well as the correlation between halo angular momentum and
large scale structure (e.g. Bailin \& Steinmetz 2005, Bett \etal
2010). One of the conclusions was that the correlation between spin
and mass (almost absent at $z=0$) seems to increase with increasing
redshift. Such a behavior could have important influences in the
modeling of properties of galaxies at high redshift.

Although it has been shown that the inclusion of baryonic physics may
affect the properties of the dark matter distribution (e.g. Gnedin
\etal 2004, Kazantzidis \etal 2004, Libeskind \etal 2010, Knebe \etal
2010, Bett \etal 2010), it also known that the strength of this effect
strongly depends on the implemented baryonic physics (Duffy \etal
2010). The aim of this work is to extend the findings of M08 to higher
redshift and to provide a comprehensive study of the evolution and
correlation of the properties of DM haloes from $z=2$ to the present
day. Such a study is a basic ingredient to understand the complex
problem of galaxy formation and evolution.

Our paper is organized as follows. In Section \ref{sec:sims} we
describe the set of simulations we used, present definitions and
describe the numerical procedures.  In Section \ref{C-Mrelation} we
present our results for the mass and redshift dependence of the
concentration parameter, present a novel and physically motivated
approach to understand the time evolution of the concentration
parameter following arguments based on the spherical collapse
model. Next we discuss our results for the mass and redshift
dependence of the shape (Section \ref{sec:shape}) and spin parameter
(Section \ref{sec:spin}).  Finally in Section \ref{sec:summ} we
summarize our findings and discuss on our results and their influence
on the modeling of galaxy properties.


\section{Numerical Simulations} 
\label{sec:sims}

All simulations have been performed with {\sc pkdgrav}, a tree code
written by Joachim Stadel and Thomas Quinn (Stadel 2001). The code
uses spline kernel softening, for which the forces become completely
Newtonian at 2 softening lengths.  Individual time steps for each
particle are chosen proportional to the square root of the softening
length, $\epsilon$, over the acceleration, $a$: $\Delta t_i =
\eta\sqrt{\epsilon/a_i}$. Throughout, we set $\eta = 0.2$, and we keep
the value of the softening length constant in comoving coordinates
during each run. The physical values of $\epsilon$ at $z=0$ are listed
in Table \ref{tab:sims}.  Forces are computed using terms up to
hexadecapole order and a node-opening angle $\theta$ which we change
from $0.55$ initially to $0.7$ at $z=2$.  This allows a higher force
accuracy when the mass distribution is nearly smooth and the relative
force errors can be large.  The initial conditions are generated with
the {\sc grafic2} package (Bertschinger 2001).  The starting redshifts
$z_i$ are set to the time when the standard deviation of the smallest
density fluctuations resolved within the simulation box reaches $0.2$
(the smallest scale resolved within the initial conditions is defined
as twice the intra-particle distance).


\begin{table}
\begin{center}
\begin{tabular}{|c|c|c|c|c|c|}\hline \hline

Name  & Box Size & N & $m_i$ & $\epsilon$ & $N_{min}> 500$  \\
      &          &   &       &            &  z=0,2          \\ \hline \hline

B20   & 14.4   & $250^3$  & 1.37e7  & 0.43 & 974,   1006  \\
B30   & 21.6   & $300^3$  & 2.68e7  & 0.64 & 1515,  1399  \\
B40   & 28.8   & $250^3$  & 1.10e8  & 0.85 & 1119,  993   \\
B90   & 64.8   & $600^3$  & 9.04e7  & 0.85 & 13587, 12177 \\
B180  & 129.6  & $300^3$  & 5.78e9  & 3.83 & 2300,  510  \\
B300  & 216.0  & $400^3$  & 1.13e10 & 4.74 & 5840,  707  \\
$\rm{B300_2}$  & 216.0  & $400^3$  & 1.13e10 & 4.74 & 5720, 766 \\
\hline
\hline
\end{tabular}
\end{center}
\caption{Table of simulations used in this work. Note that the name of
  the simulation is related to the box size in units of Mpc. $N$
  represents the number of total particles in the box. $\epsilon$
  represents the force softening length in units of kpc $h^{-1}$ and
  the last column gives the number of haloes with more than 500
  particles at $z=0$ and $z=2$.  Masses of particles are in units of
  \hMsun\ and box sizes in units of Mpc $h^{-1}$, with $h=0.72$.}
\label{tab:sims}
\end{table}


Table \ref{tab:sims} lists all of the simulations used in this work.
We have run simulations for several different box sizes, which allows
us to probe halo masses covering the entire range $10^{10}$ \hMsun $<
M < 10^{15} h^{-1}\,\rm M_{\odot}$.  In addition, in some cases we
have run multiple simulations for the same cosmology and box size, in
order to test for the impact of cosmic variance (and to increase the
final number of dark matter haloes).

In all of the simulations, dark matter haloes are identified using a
spherical overdensity (SO) algorithm. We use a time varying virial
density contrast determined using the fitting formula presented in
Bryan and Norman (1998). We include in the halo catalogue all the
haloes with more than 500 particles inside the virial radius ($N_{\rm
  vir}>500$).

We have set the cosmological parameters according to the fifth-year
results of the Wilkinson Microwave Anisotropy Probe mission (WMAP5,
Komatsu \etal 2009), namely, $\Omega_m = 0.258$, $\Omega_L = 0.742$,
$n=0.963$, $h = 0.72$, and $\sigma_8 = 0.796$, where $\Omega_m$ and
$\Omega_L$ are the values of the density parameters at z=0.

\subsection{Halo parameters}
\label{ssec:param}

For each SO halo in our sample we determine a set of parameters,
including the virial mass and radius, the concentration parameter, the
angular momentum, the spin parameter and axis ratios (shape).  Below
we briefly describe how these parameters are defined and determined. A
more detailed discussion can be found in Macci\`o \etal (2007, 2008).

\subsubsection{Concentration parameter}
\label{ssec:c}
To compute the concentration of a halo we first determine its density
profile.  The halo centre is defined as the location of the most bound
halo particle (we define the most bound particle as the particle with
the lowest potential energy, no care about binding energy is taken
here), and we compute the density ($\rho_i$) in 50 spherical shells,
spaced equally in logarithmic radius.  Errors on the density are
computed from the Poisson noise due to the finite number of particles
in each mass shell.  The resulting density profile is fitted with a
NFW profile:

\begin{equation}
  \frac{\rho(r)}{\rhocrit} = \frac{\delta_{\rm c}}{(r/\rs)(1+r/\rs)^2},
  \label{eq:nfw}
\end{equation}

During the fitting procedure we treat both $\rs$ and $\delta_c$ as free
parameters.  Their values, and associated uncertainties, are obtained via a
$\chi^2$ minimization procedure using the Levenberg \& Marquardt method.  We
define the r.m.s. of the fit as:
\begin{equation}
\rhorms = \frac{1}{N}\sum_i^N { (\ln \rho_i - \ln \rho_{\rm m})^2}
\label{eq:rms}
\end{equation}
where $\rho_{\rm m}$ is the fitted NFW density
distribution.\footnote{A more conservative notation for the r.m.s of
  the fit would be $\sigma^2_{\rho}$. Nevertheless we keep the use of
  $\rhorms$ to be consistent with the notation used in Macci\`o \etal
  (2007, 2008)}
Finally, we define the concentration of the halo, $\cvir
\equiv\Rvir/\rs$, using the virial radius obtained from the SO
algorithm, and we define the error on $\log c$ as
$(\sigma_{\rs}/\rs)/\ln(10)$, where $\sigma_{\rs}$ is the fitting
uncertainty on $\rs$.

\subsubsection{Spin parameter}
\label{sec:spinpar}

The  spin  parameter is  a  dimensionless  measure  of the  amount  of
rotation of a  dark matter halo.  We use  the definition introduced by
Bullock \etal (2001b):
\begin{equation}
\lambda'=\frac{\Jvir}{\sqrt{2}\Mvir\Vvir\Rvir}
\end{equation}
where $\Mvir$ is the mass interior to $\Rvir$, $\Jvir$ is the total
angular momentum of that mass distribution and $\Vvir$ is its circular
velocity at the virial radius.  See Macci\`o \etal (2007) for a
detailed discussion and for a comparison of the different definitions
of the spin parameter.

\subsubsection{Shape parameter}
\label{sec:Defshapes}
Determining the shape of a three-dimensional distribution of particles is a
non-trivial task (e.g., Jing \& Suto 2002).  Following Allgood \etal (2006),
we determine the shapes of our haloes starting from the inertia tensor.  As a
first step, we compute the halo's $3 \times 3$ inertia tensor using all the
particles within the virial radius.  Next, we diagonalize the inertia tensor
and rotate the particle distribution according to the eigenvectors.  In this
new frame (in which the moment of inertia tensor is diagonal) the ratios
$a_3/a_1$ and $a_3/a_2$ (where $a_1 \geq a_2 \geq a_3$) are given by:
\begin{equation}
{a_3 \over a_1} = \sqrt{ { \sum m_i z_i^2} \over \sum { m_i x_i^2}}\\
{a_3 \over a_2} = \sqrt{ { \sum m_i z_i^2} \over \sum { m_i y_i^2}}.
\end{equation}

Next we again compute the inertia tensor, but this time only using the
particles inside the ellipsoid defined by $a_1$, $a_2$, and $a_3$.
When deforming the ellipsoidal volume of the halo, we keep the longest
axis ($a_1$) equal to the original radius of the spherical volume
($\Rvir$). We iterate this procedure until we converge to a stable set
of axis ratios. Although this iterative procedure can indeed change
the mass contained inside of the ellipsoid, we checked that variations
are nevertheless below 20\%. We will therefore not consider those
changes in mass when showing mass-shape relations, and we will work
always with virial masses.

\subsection{Relaxed - Unrelaxed  haloes}
\label{sub:relax}

Our halo finder (and halo finders in general) does not distinguish
between relaxed and unrelaxed haloes.  There are many reasons why we
might want to remove unrelaxed haloes.  First and foremost, unrelaxed
haloes often have poorly defined centers, which makes the
determination of a radial density profile, and hence of the
concentration parameter, an ill-defined problem.  Moreover unrelaxed
haloes often have shapes that are not adequately described by an
ellipsoid, making our shape parameters ill-defined as well.

Following Macci\`o \etal (2007), we decide to use a combination of two
different parameters $\rhorms$ and $\xoff$ to determine the dynamical
status of a given dark matter halo. The first quantity $\rhorms$ is
defined as the r.m.s.  of the NFW fit to the density profile
(performed to compute $c_{vir}$).  While it is true that $\rhorms$ is
typically high for unrelaxed haloes, haloes with relatively few
particles also have a high $\rhorms$ (due to Poisson noise) even when
they are relaxed; furthermore, since the spherical averaging used to
compute the density profiles has a smoothing effect, not all unrelaxed
haloes have a high $\rhorms$.  In order to circumvent these problems,
we combine the value of $\rhorms$ with the $\xoff$ parameter, defined
as the distance between the most bound particle (used as the center
for the density profile) and the center of mass of the halo, in units
of the virial radius.  This offset is a measure for the extent to
which the halo is relaxed: relaxed haloes in equilibrium will have a
smooth, radially symmetric density distribution, and thus an offset
that is virtually equal to zero.  Unrelaxed haloes, such as those that
have only recently experienced a major merger, are likely to reveal a
strongly asymmetric mass distribution, and thus a relatively large
$\xoff$. Although some unrelaxed haloes may have a small $\xoff$, the
advantage of this parameter over, for example, the actual virial
ratio, $2T/V$, as a function of radius (e.g. Macci\`o, Murante \&
Bonometto 2003), is that the former is trivial to evaluate.  Following
Macci\`o \etal (2007), we split our halo sample into unrelaxed and
relaxed haloes.  The latter are defined as the haloes with $\rhorms <
0.5$ and $\xoff < 0.07$.  About 70\% of the haloes in our sample
qualify as relaxed haloes at $z=0$.

To check for the effect of changing the definition of relaxed haloes,
we have computed the median concentration (as shown in the next
section) using different values of $\xoff$. Changing the value of this
parameter by 25\% (above and below 0.07) induces changes no larger
than 5\% in the median concentration of dark matter haloes. We
conclude that choosing $\xoff = 0.07$ our results are robust enough
against variations in the definition of relaxed population of
haloes. In what follows we will just present results for haloes which
qualify as relaxed.

\section{Concentration: Mass and Redshift dependence}
\label{C-Mrelation}

\begin{figure}
  \includegraphics[width=8.0cm,angle=270]{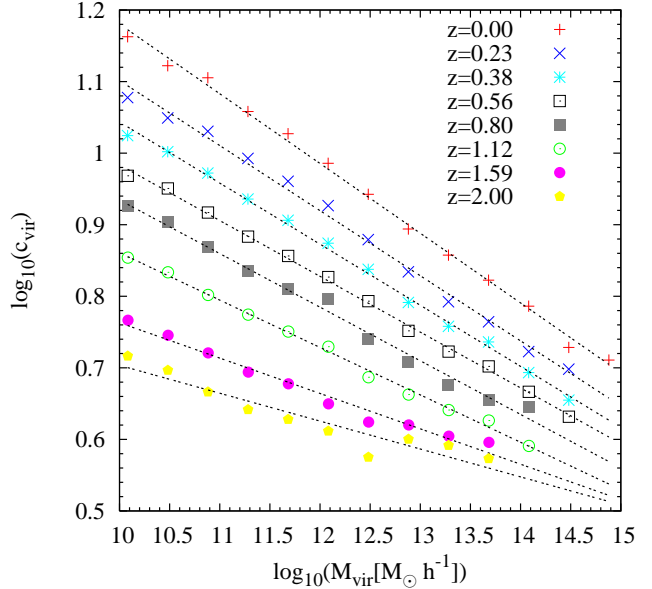}
  \caption{Mass and redshift dependence of the concentration
    parameter. The points show the median of the concentration as
    computed from the simulations, averaged for each mass bin. Lines
    show their respective linear fitting to eq. \ref{eq:Cfit}.}
  \label{fig:Cwithfit}
\end{figure}

In figure \ref{fig:Cwithfit} we show the median \cM relation for
relaxed haloes in our sample at different redshifts.  Haloes have been
binned in mass bins of 0.4 dex width, the median concentration in each
bin has been computed taking into account the error associated to the
concentration value (see \ref{ssec:c}, and M08).  In our mass range
the \cM relation is well fitted by a single power law at almost all
redshifts. Only for $z=2$ we see an indication that the linearity of
the relation in log space seems to break, in agreement with recent
findings by Klypin \etal (2010).

The best fitting power law can be written as:
\begin{equation}
\log(c) = a(z)\log(M_\mathrm{vir}/[\hMsun]) + b(z)
\label{eq:Cfit}
\end{equation}

\begin{figure}
  \includegraphics[width=8.0cm,angle=270]{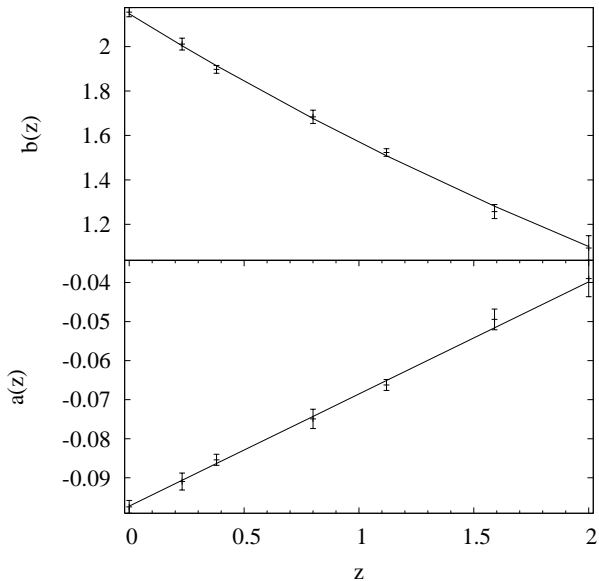}
  \caption{Redshift evolution of the parameters $a(z)$ and $b(z)$ of
    the linear fitting of the \cM relation. Points represent the
    result of the fitting shown in figure \ref{fig:Cwithfit} for each
    redshift while lines are the fit to the functions in eq:.
    \ref{Pfita} and \ref{Pfitb}. Error bars show the error obtained
    from the fitting to eq. \ref{eq:Cfit}.}
  \label{fig:Cparams}
\end{figure}

The fitting parameters $a(z)$ and $b(z)$ are functions of redshift as
shown in figure \ref{fig:Cparams}. The evolution of $a$ and $b$ can be
itself fitted with two simple formulas that allow to reconstruct the
\cM relation at any redshifts:
\begin{equation}
a(z) = wz  - m
\label{Pfita}
\end{equation}
\begin{equation}
b(z) = \frac{\alpha}{(z+\gamma)}+\frac{\beta}{(z+\gamma)^2}
\label{Pfitb}
\end{equation}

\begin{figure}
  \includegraphics[width=8.0cm,angle=270]{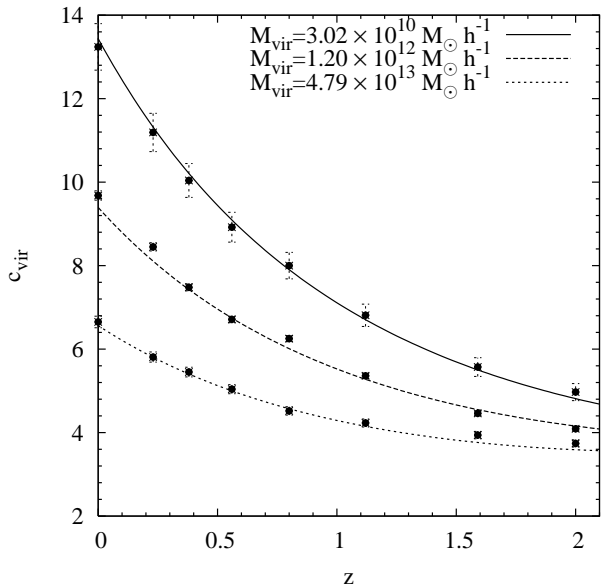}
  \caption{Reconstruction of the \cM relation for different mass bins
    as a function of redshift as computed from the model fitting
    function, equations \ref{eq:Cfit} to \ref{Pfitb}. Points show the
    data of our simulations while the lines show the result of the
    reconstruction. Error bars show the error in the median
    concentration for each redshift.}
  \label{fig:Creconst}
\end{figure}

\noindent
Where the additional (constant) fitting parameters have been set equal
to: $w=0.029$, $m=0.097$, $\alpha=-110.001$, $\beta=2469.720$ and
$\gamma=16.885$.  Figure \ref{fig:Creconst} shows the reconstruction
of the \cM relation for different mass bins as a function of redshift
using the approach described above.  It shows that our (double)
fitting formulas are able to recover the original values of the halo
concentration with a precision of 5\%, for the whole range of masses
and redshifts inspected. It has been shown by Trenti \etal (2010) that
using $\Nvir$ between 100 and 400 particles is enough to get good
estimates for the properties of halos, nevertheless in order to look
for systematics we re-computed $\cvir$ varying the minimum number of
particle inside $R_{\rm vir}$, using 200, 500 and 1000 particles. No
appreciable differences (less than 2\%) were found in our results for
the median. We also checked that our results do not change notably by
changing the definition of ``relaxed'' haloes (i.e. changing the cut
in $\xoff$ or $\rhorms$).

It is interesting to compare our results with M08, which shares some
of the simulations presented in this work. Our results for the \cM
relation at $z=0$ are slightly different to those presented in M08: we
found $a=-0.097$ and $b=2.155$, while M08 found $a=-0.094$ and
$b=2.099$.  The difference is less than 3\%, and it is mainly due to
low mass haloes.  In this work we included three new simulations, B30,
B90 and B$300_2$. Two of these (B30 and B90) increase the statistics
of our halo catalogs at the low mass end, providing a better
determination of the \cM relation for $M \approx 10^{11} h^{-1}\,\rm
M_{\odot}$.  We are confident that the inclusion those new simulations
led to an improvement over the results of previous works.

\subsection{Understanding the concentration  evolution}
\label{ssec:under}

We want now to explore in more detail the physical mechanism driving
the mass and redshift dependence of the concentration parameter, this
understanding could take us to a better interpretation of the time
evolution of the dark matter density profile. As $\cvir$ is defined as
the ratio between $\Rvir$ and $\rs$ we will look at the time evolution
of those quantities for different halo masses, and try to see if we
can extract some physical insights about the evolution of $\cvir$.

Since the evolution of the concentration parameter is strongly
correlated to the mass growth history of the dark matter halo
(e.g. Wechsler \etal 2003; Zhao \etal 2009) we need to construct
merger trees for haloes in our simulation boxes.  Here we briefly
describe the procedure used to build merger trees (for more details
see Neistein, Macci\`o \& Dekel 2010).

We link haloes $A$ and $B$, with particle number $N_A$ and $N_B$, in
two consecutive snapshots at redshift $z_A$ and $z_B$ ($z_A < z_B$),
if they success to satisfy a list of requirements:

\begin{itemize}
\item if $N_B < N_A$, they have at least $0.5N_B$ particles in common
\item if $N_A < N_B$, they have at least $0.5N_A$ particles in common
\item Halo $B$ does not contribute to any other halo in $z_A$ with more
  particles than it does for halo $A$
\end{itemize}

We will assume that the evolution of the structural properties of the
halo is traced by the most massive progenitor, therefore we will
present results only for the main branch of the tree.  We follow along
the tree the evolution of $\Mvir$, $\Rvir$, $\rs$ and $\cvir$. Since
individual histories of haloes can be very different we focus our
study on the time evolution of the mean value of these quantities.
These averages are computed binning the histories by mass at $z=0$
with each mass bin having width of 0.4 dex. One may argue that
following only the merger tree main branch of haloes indentified at
z=0 could introduce a bias when comparing to the evolution of haloes
at earlier cosmic times. In order to verify the stability of our
results we rebuilt the merger histories for all haloes starting from
z=0.5 and z=1.0 (i.e. without taking into account the future evolution
of an halo, like, for example, being or not in the final catalogue at
z=0).  We found no differences in the evolution of the mean values for
$\Rvir$ or $\Mvir$ with respect to the original merger tree built from
z=0.

For the averages we include only histories with the same length, that
is, histories that in the entire box evolve from the same $z_{\rm
  init}$ to $z=0$. $z_{init}$ was chosen by numerical-statistic
reasons. As we are computing the mean on several properties of haloes
in the merger histories along the time, we wanted to make sure to have
at each redshift a statistically meaningful number of histories. The
extension in redshift is then constrained by the resolution of our
simulation and the number of histories we choose for our study. At the
end we found $z_{init} \sim 3.5$ to be a good compromise.  Moreover we
only use histories in which the fitting of the density profile was
successful in the majority of the snapshots.  Any halo was allowed to
be ``unrelaxed'' for a maximum of 2 consecutive snapshots. In this
case, we used as criterion for the goodness of the density profile fit
$\rho_{\rm rms} > 0.5$.  Using both our criteria of relaxation
($\rho_{\rm rms}$ and $\xoff$) the number of histories becomes too
small to be statistically meaningful.  To be able to follow the
histories longer in redshift we reduced the minimum number of
particles per halo (inside the virial radius) from 500 (as used in the
previous section) to 200. We verified that using 200 particles inside
the virial radius the fitting of the density profile is still
acceptable for the purposes of this section.

Finally, since we are constraining the mass histories in redshift
extent, we cannot overlap data among different boxes, then we have
computed average quantities for individual boxes independently.  In
the following we will present results coming from the B90 box, other
boxes show very similar behaviours. Applying all of this selection
criteria on B90 results in a set of 2300 histories with the different
mass bins having between 35 and 800 halos each.

\begin{figure}
  \includegraphics[width=8.0cm,angle=270]{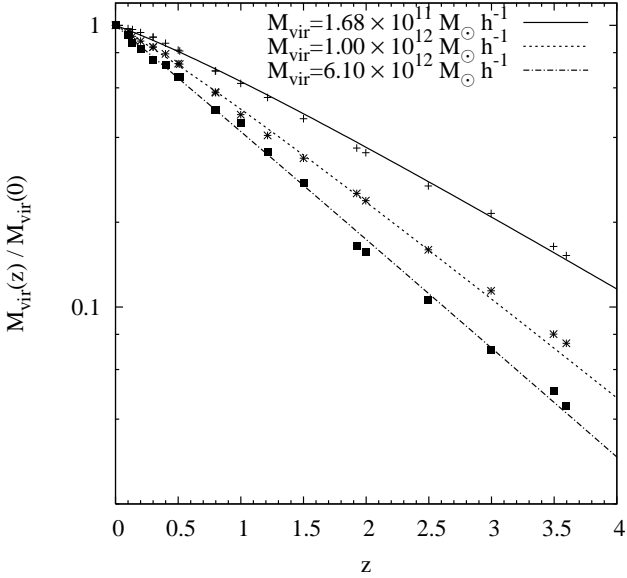}
  \caption{ Mean mass accretion history for haloes in the 90 Mpc box
    computed for different mass bins according to their mass at z=0
    (from $1.68 \times 10^{11}$ to $2.8 \times 10^{13.5}$ \hMsun with
    every mass bin of width 0.4 dex) as described in the main
    text. Here is evident the different shape of the mass growth
    history of haloes of different masses.  Lines show the best fit to
    Eq \ref{eq:Mhist} which shows to be a good description for all of
    our mass histories. The values of ($\gamma,\beta$) for the fit are
    (0.649,0.273), (0.865,0.259) and (0.900,0.045) for each line (low
    to high mass) respectively.}
  \label{fig:Mhist}
\end{figure}

In figure \ref{fig:Mhist} we show the average mass accretion history
(normalized to the $z=0$ value) for dark matter haloes in three
different mass bins centered on: $1.68\times10^{11} h^{-1}\,\rm
M_{\odot}$, $1.0\times10^{12}$\hMsun and $6.1\times10^{12} h^{-1}\,\rm
M_{\odot}$. The data are well described by the two parameter function
(McBride \etal 2009)

\begin{equation}
  M(z) = M_0(1+z)^\beta\exp(-\gamma z)
  \label{eq:Mhist}
\end{equation}

Here $M_0$ is the mass at $z=0$ and $\beta$ and $\gamma$ are free
parameters related to the mass growth rate at low $z$.  Note that
although for most haloes $\beta\neq0$, when $\beta=0$ the profile
assumes the exponential shape adopted by Wechsler \etal (2003) with
$\gamma=\ln(2)/z_f$.

\begin{figure}
  \includegraphics[width=8.0cm,angle=270]{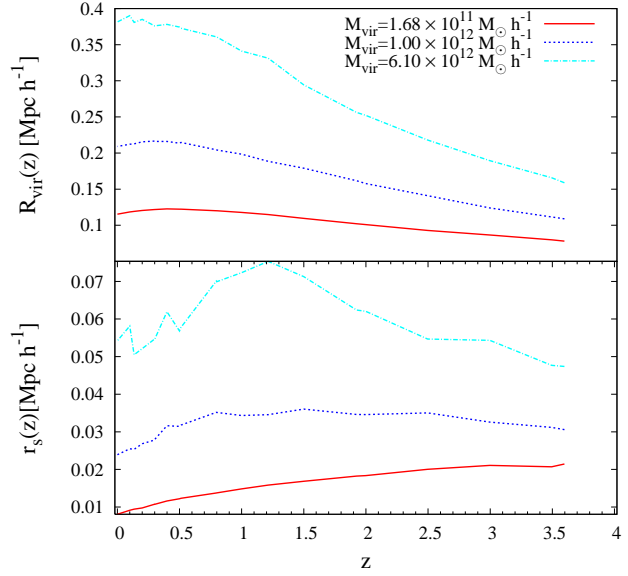}
  \caption{(Top) Time evolution of the virial radius as computed from
    the averaged merger histories for haloes with final mass of
    $1.68\times10^{11}$, $1.0\times10^{12}$ and $6.1\times10^{12}
    h^{-1}\,\rm M_{\odot}$. (Bottom) Time evolution of the averaged
    scale length $\rs$ of the dark matter haloes in the same mass
    bins. The mass binning is the same as in figure \ref{fig:Mhist}.}
  \label{fig:Rvir}
\end{figure}

In figure \ref{fig:Rvir} (top panel) we show the redshift evolution of
the virial radius for haloes with final masses: $1.68\times10^{11}
h^{-1}\,\rm M_{\odot}$, $1.0\times10^{12}$\hMsun and $6.1\times10^{12}
h^{-1}\,\rm M_{\odot}$.  For all mass scales the virial radius grows
with decreasing redshift reaching a maximum, and then starts a slow
decrease. The redshift at which this maximum is reached depends on the
mass of the halo, lower mass haloes reach that maximum earlier than
more massive ones.  Given the definition of $R_{\rm vir}$:
\begin{equation}
R_{\rm vir}(z) = \left(\frac{3M_{\rm vir}(z)}{4\pi\Delta_{\rm vir}(z)\rho_{\rm c}(z)}\right)^{1/3}
\label{eq:Rvir}
\end{equation}
\noindent
its time evolution can be understood as follows. 
At high redshift the radius of the virialized region $\Rvir$ grows due
to the growth of the halo mass $M(z)$ but is also subject to the
effects of the cosmological background via $[\Delta_{\rm
    vir}(z)\rho_{\rm c}(z)]^{-1/3}$, which in this case is a slowly
decreasing function with decreasing redshift. At high redshift the
growth of $M(z)$ dominates, forcing $\Rvir$ to grow, at low redshift
the growth rate of the halo mass becomes weaker compared to the
decrease of the factor $[\Delta_{\rm vir}(z)\rho_{\rm c}(z)]^{-1/3}$,
slowing the growth of $\Rvir$. This behaviour depends on the halo
mass: at different redshifts haloes of different masses grow at
different rates (see figure \ref{fig:Mhist}), on the other hand, for a
given redshift, the factor $[\Delta_{\rm vir}(z)\rho_{\rm
    c}(z)]^{-1/3}$ is mass independent. As a result the point at which
$\Rvir$ reaches its maximum happens later in time for massive haloes
than for low mass ones.

The bottom panel of figure \ref{fig:Rvir} shows the redshift evolution
of the halo scale length $\rs$, which has a trend similar to $\Rvir$:
it grows with time, reaches a maximum and then starts to
decrease. Once again the redshift at which this maximum is achieved
depends on the mass of the halo, with low mass haloes reaching the
maximum at earlier times.

The behaviour of $\rs$ (and $\Rvir$) is strongly reminiscent of the
behaviour of a perturbation in the spherical collapse model. It seems
to suggest that the inner region of an halo (within $\rs$) evolves in
a decoupled way compared to the global perturbation (within $\Rvir$).
It is then possible to model the inner region as a perturbation of
density $\rho_{\rm s}$ (defined as the density inside $\rs$) that
evolves within the background density $\rho_{\rm vir}(z)= \Delta_{\rm
  vir}(z) \rho_{\rm c} (z)$.

In analogy with the spherical collapse model we want to look for the
evolution of the density contrast of this perturbation: $\Delta_{\rm
  s}(z) = \rho_{\rm s} (z)/\rho_{\rm vir} (z)$.  This inner density
contrast is well described by the following formula:
\begin{equation}
  \Delta_{\rm s} (z)= \frac{A}{z+\epsilon(M)}
  \label{eq:rsth}
\end{equation}
where $A=50$ and $\epsilon(M) = 0.3975\log(M_{\rm vir}(z=0)/[\hMsun])-
4.312$.  Equation \ref{eq:rsth} implies that: i) $\rho_{\rm s} >
\rho_{\rm vir}$ at all redshifts, ii) $\Delta_{\rm s}$ is a growing
function of the redshift, implying a faster growth of the inner
density with respect to the mean density of the halo and iii)
$\Delta_{\rm s}$ depends on the final mass of the halo, and it has
larger values for high mass haloes.

It is now possible to interpret the evolution of the \cM relation in
the light of our findings. Figure \ref{fig:Creconst} shows that the
growth rate of the concentration depends on the halo mass, with low
mass haloes experiencing a faster concentration evolution. Moreover,
at fixed mass, the evolution of the \cM relation is faster at lower
redshifts.  As can be seen from figure \ref{fig:Rvir}, at early times
$\rs$ grows simultaneously with $\Rvir$, then, with decreasing
redshift, the growth of $\rs$ slows down, reaches a maximum, and
starts to decrease. When $\Rvir$ grows together with $\rs$, the
concentration of the halo stays approximately constant or slightly
increases. Then when $\rs$ slows down its growth rate and starts to
``contract'' the concentration of the halo grows rapidly. In the case
of high mass haloes the time at which the dynamics of the inner region
decouples from the outermost part, happens at later times compared to
low mass haloes. As a consequence more massive haloes have a more
extended period in which the concentration is a slowly growing
function of time as shown in figure \ref{fig:Creconst}.  Only in
recent cosmic times, due to the late collapse of the inner part the
concentration starts to grow at a higher rate also for larger masses.

A crucial point in this process is the moment when the inner part of
the halo decouples from the outer one. This specific time is strongly
mass dependent with smaller structures (i.e. the more non linear ones)
decoupling earlier.  This effect explains the change in the slope of
the \cM relation, as described in figure \ref{fig:Cparams} by the
evolution of $a(z)$, and also explains why a simple scaling of the
redshift zero \cM relation with a factor of the form $(1+z)^\alpha$
(e.g. Bullock \etal 2001a) is not able to reproduce the simulated
data.

\section{Halo shapes}
\label{sec:shape}

\begin{figure}
  \includegraphics[width=8.0cm,angle=270]{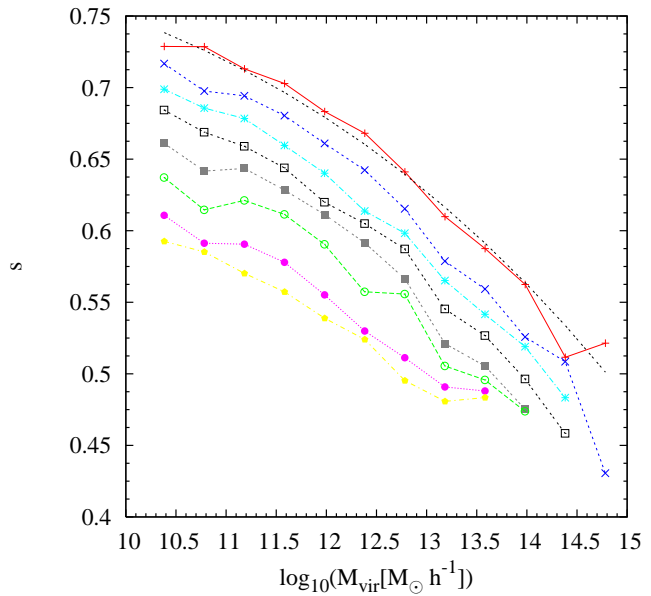}
  \caption{Mass and time evolution of the shape of dark matter haloes
    quantified via the $s\equiv a_3/a_1$ ratio. Points represent our
    data while the dashed line shows the fitting to equation
    \ref{eq:NewShape} to $z=0$. Similar results for the fitting are
    obtained for $z>0$ and for clarity are not shown in the plot. The
    color code is the same used in figure \ref{fig:Cwithfit} and shows
    different redshifts. Values of the fit parameters for all
    redshifts can be found in the appendix.}
  \label{baz}
\end{figure}

\begin{figure}
  \includegraphics[width=8.0cm,angle=270]{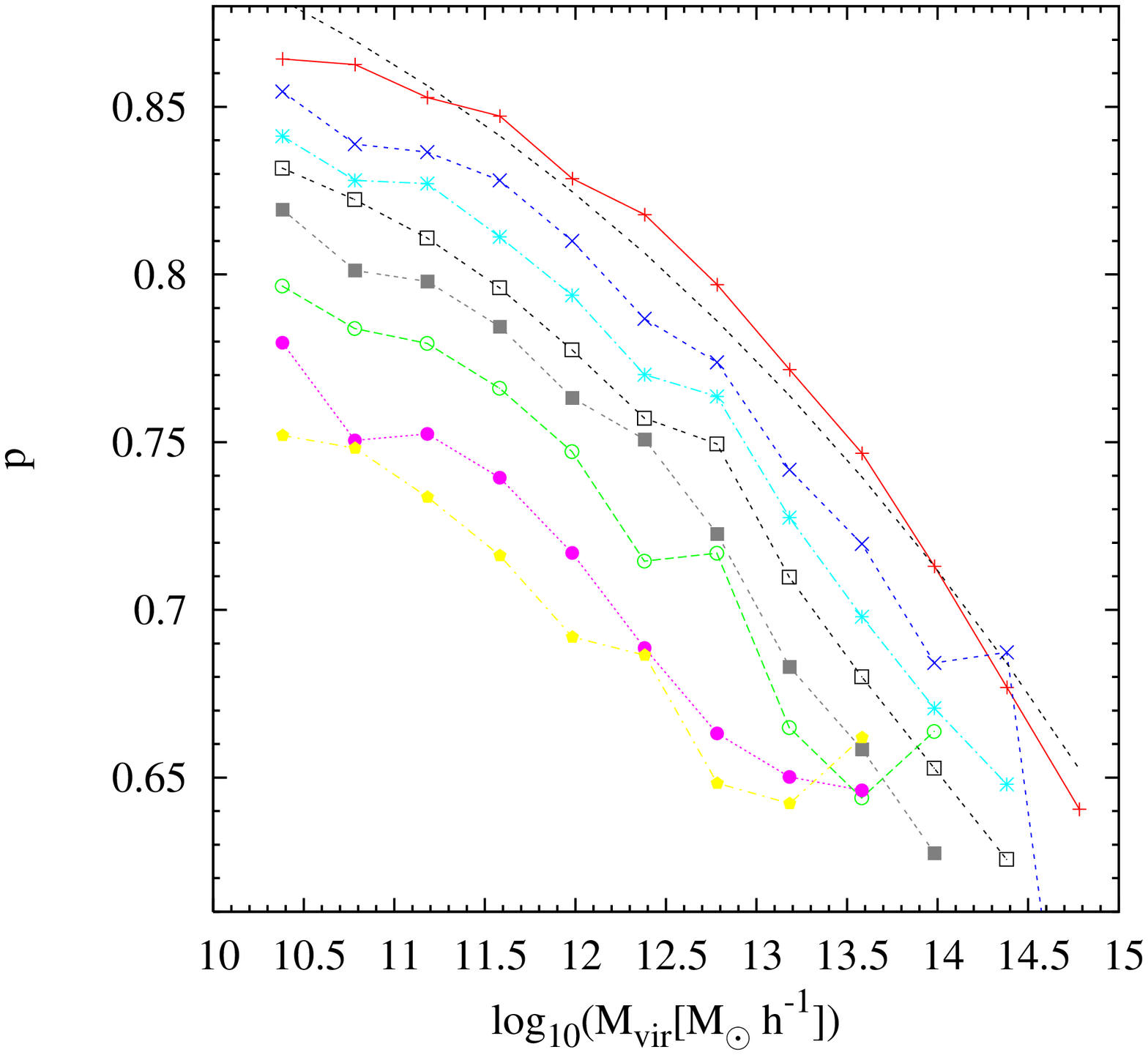}
  \caption{Mass and time evolution of the shape of dark matter haloes
    quantified via the $p\equiv a_3/a_2$ ratio. Points represent our
    data while the dashed line shows the fitting to equation
    \ref{eq:NewShape} to $z=0$.  Similar results for the fitting are
    obtained for $z>0$ and for clarity are not shown in the plot. The
    color code is the same used in figure \ref{fig:Cwithfit} and shows
    different redshifts. Values of the fit parameters for all
    redshifts can be found in the appendix.}
  \label{caz}
\end{figure}

We now turn our attention to the evolution of halo shape with redshift
and its dependence on halo mass. As was described in section
\ref{sec:Defshapes}, for each dark matter halo we compute the axis
ratios $s=a_3/a_1$ and $p=a_3/a_2$, where $a_1 \geq a_2 \geq a_3$ are
the major, intermediate and minor axis of the halo mass
distribution. It is worth to note that in our notation using only $s$
and $p$ is enough to determine the shape of a dark matter halo. The
condition for a halo to be oblate will be $s \sim p < 1$, while the
condition to be prolate will be $s<p$ with $p \sim 1$, where the
obvious condition for sphericity is $s \sim p \sim 1$ and triaxiality
will be any other not filling any of the previous requirements. Figure
\ref{baz} shows the evolution of the shape parameter $s$ as a function
of redshift and mass. Figure \ref{caz} is the analogue of figure
\ref{baz} but for the $p$ parameter, only results for haloes with
$N_{\rm vir}>1000$ are shown; after several convergence tests we found
that this number ensures both numerical stability of the axis
determination and a fairly large statistical sample.  Both figures
show that on average halos are preferentially triaxial, where the most
massive haloes tend to be the most ellipsoidal ones, while lower mass
haloes tend to be closer to spherical , and this trend seems to be
redshift independent.  This seems to suggest a simple scenario where
the most massive haloes are the more extended ones (and less
concentrated) and hence more strongly affected by tidal torques.  For
each redshift we fit the data with a power law of the form of eq:
\ref{eq:Cfit}. The resulting parameters of this fit are listed in the
Appendix.

\begin{figure}
  \includegraphics[width=8.0cm,angle=270]{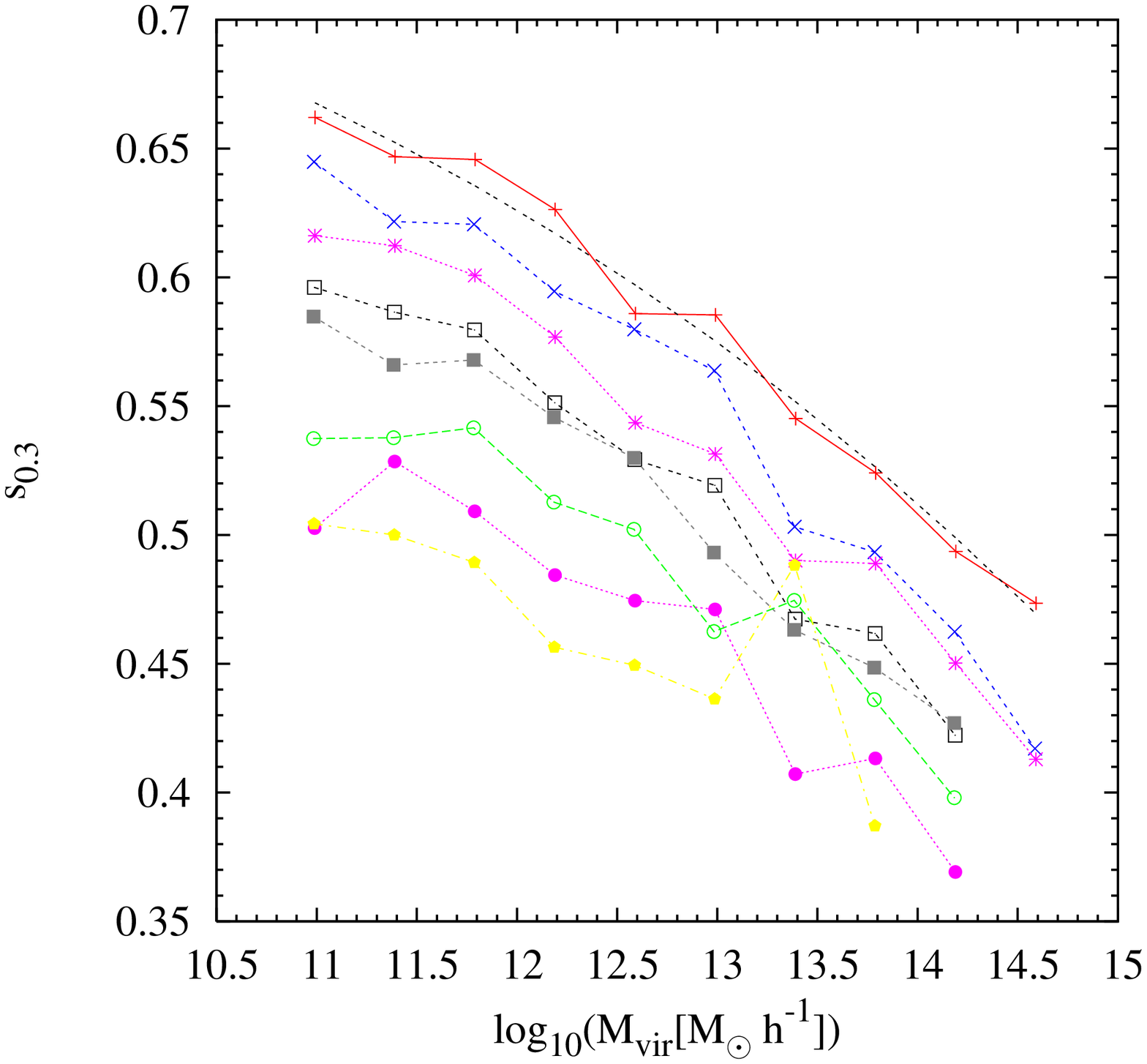}
  \caption{Mass and time evolution of the shape of dark matter haloes
    quantified via the $s_{0.3}\equiv a_3/a_1$ ratio measured at
      $0.3\Rvir$. Points represent our data while the dashed line
    shows the fitting to equation \ref{eq:NewShape} to $z=0$.  Similar
    results for the fitting are obtained for $z>0$ and for clarity are
    not shown in the plot. The color code is the same used in figure
    \ref{fig:Cwithfit} and shows different redshifts. Values of the
    fit parameters for all redshifts can be found in the appendix.}
  \label{ca3z}
\end{figure}

In order to make a more direct comparison with the results of Allgood
\etal (2006) we also computed the shape parameter $s$ using only
particles within the inner 30\% of the virial radius $s_{0.3}$ for
haloes with $N_{\rm vir}>4000$. The results are shown in figure
\ref{ca3z}. {Allgood \etal (2006) proposed a fitting formula for the
  shape parameter given by},

\begin{equation}
  s(z,M) = \alpha \left(\frac{\rm M_{vir}}{M_*(z)} \right)^{\beta}
\label{eq:fshape}
\end{equation}

\noindent
where $M_*(z)$ is the characteristic non-linear mass at $z$ such that
the rms top-hat smoothed overdensity at scale $\sigma(M_*,z)$ is
$\delta_c=1.68$, and $\alpha$ and $\beta$ are free parameters. The
quantity $M_*(z)$ should in principle contain all the information
about the cosmological model, making the other fitting parameters
cosmological independent.

As already shown by M08 and Bett \etal (2007) (at $z=0$),
Eq. (\ref{eq:fshape}) does not provide a good fit to the data.  In
order to get a reasonable fit we need to add an explicit redshift
dependence to the parameters $\alpha$ and $\beta$, while in original
the model proposed by Allgood \etal, the redshift dependence of the
shape parameter $s$ was described by the quantity $M_*(z)$.  A second
drawback of Eq. (\ref{eq:fshape}) can be clearly seen in figures
\ref{baz}, \ref{caz} and \ref{ca3z}. The convex-curved shape of the
data deviates from a simple power-law behavior.  In order to better
model the data we modified Eq. (\ref{eq:fshape}) into
\begin{equation}
s(z,M) = \alpha (\log(\rm M_{vir}/[\hMsun])^{4} + \beta.
\label{eq:NewShape}
\end{equation}
The values of the new fitting parameters $\alpha$ and
$\beta$ are reported in table \ref{tab:Table3} in the appendix.

Finally we compare the inner ($s_{0.3}$) to the outer shape ($s$); we
found that haloes are slightly aspherical in their central region
(5-10\% effect) as already noticed by Allgood \etal (2006), who
ascribed this difference to the higher number of substructure in the
central region with respect to the external one. However it should be
kept in mind that the internal region is the most likely to be
affected by baryonic physics (e.g. Kazantzidis \etal 2004) and hence
the inclusion of dissipational effects is needed in order to drive
more quantitative conclusions.

\section{Halo Spin parameter}
\label{sec:spin}

\begin{figure}
  \includegraphics[width=7.8cm,angle=270]{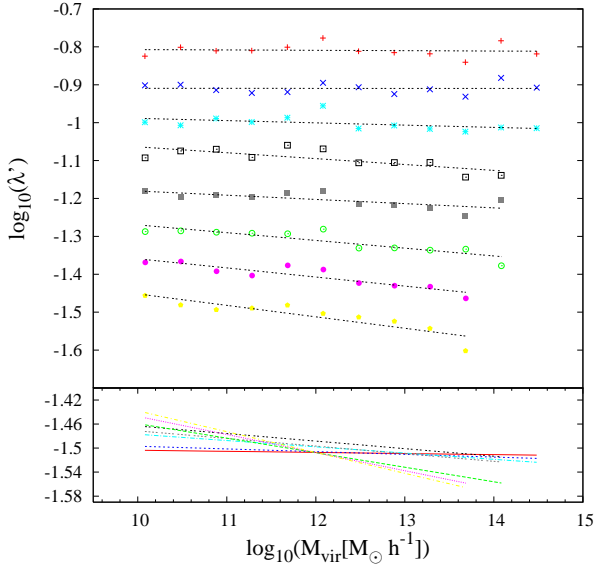}
  \caption{Mass and time evolution of the spin parameter
    $\lambda'$. (Top) Points are the median from our simulations while the solid
    lines represent the linear fitting to the data. Every line have been
    shifted by a constant factor of 0.1 from $z=2$ and all of them show
    approximately the same mean value, nevertheless as z increases, a weak
    dependence of the spin on the mass of the haloes starts to become evident.
    The color code is the same used in figure \ref{fig:Cwithfit} and shows
    different redshifts. (Bottom) Unshifted linear fitting to $\lambda'$. The
    color code of each line is the same used in the plot on top.}
  \label{spin}
\end{figure}

The relation between spin parameter ($\lambda'$) and halo mass is
shown in figure \ref{spin}, where for each redshift value we fit the
data with a power law and in analogy with eq. \ref{eq:Cfit} we denote
the gradient and intercept in log space as $a(z)$ and $b$
respectively. The parameters of these fits are shown in the Appendix.
For low redshift ($z=0; 0.23)$ we found our results to be consistent
with the spin parameter being almost mass independent, in agreement
with Macci\`o \etal (2007; 2008) and Bett \etal 2007. At higher
redshift ($z>0.5$) a mass dependence of the spin starts to develop;
with high mass haloes having on average a lower value of
$\lambda'$. This is consistent with the results of Knebe \& Power
(2008), who performed a detailed study of the spin-mass relation for
$z=1$ and $z=10$.

\begin{figure}
  \includegraphics[width=7.8cm,angle=270]{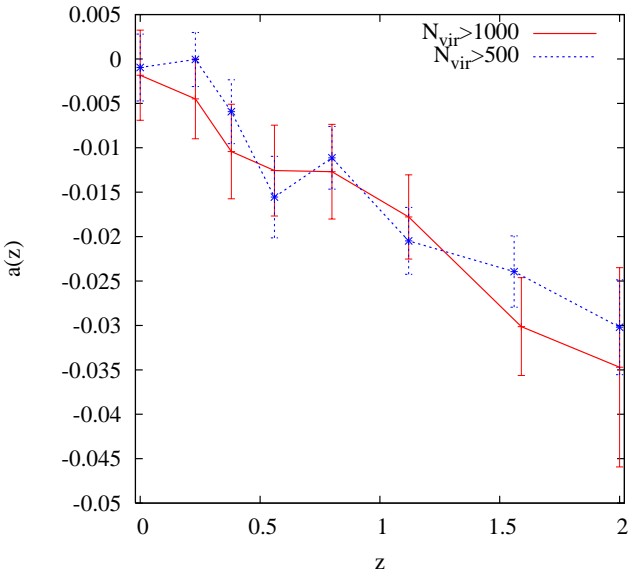}
  \caption{Time evolution of the parameters of fitting $a(z)$ and
    $b(z)$ to equation \ref{eq:Cfit} for the spin-mass relation for
    haloes with at least 500 (blue) and 1000 (red) particles inside
    the virial radius. Error bars show the error in the value of the
    parameter a given from the fitting.}
  \label{spinPars}
\end{figure}

The evolution of the slope of the spin-mass relation is more evident
in figure \ref{spinPars} where the values of the $a(z)$ parameter are
shown (see eq: \ref{eq:Cfit}).  As can be seen in the figure, $a$
evolves from being consistent with 0 at $z=0$ to negative values for
increasing redshift.  Knebe \& Power (2008) noted that this behavior
is almost insensitive to the mass binning and details of the selection
criteria for the relaxed halo population, but they found that the
number of particles per halo is an important parameter.  In order to
verify the stability of our results against the minimum number of
particles per halo, we show in figure \ref{spinPars} the redshift
evolution of the slope of the spin-mass relation for two different
choices of $N_{\rm vir}$, namely 500 and 1000 particles. As can be
seen in the figure, the change in $\Nvir$ does change (slightly) the
value of the fitted slope $a(z)$, nevertheless this change is still
within the statistical errors and does not affect the overall trend.
 
It could be possible that the mass dependence we observe could be
partially due to the mixing of different box sizes (and hence haloes
with different resolution).  For this purpose we analyzed the
spin-mass relation independently in different boxes (B90, B180 and
B300). We found no significant differences between our global analysis
and the single box results.  We can conclude that no numerical
artifacts are affecting our results and that indeed spin parameter and
halo mass are weakly correlated at high redshift.

\begin{figure}
  \includegraphics[width=8cm,angle=270]{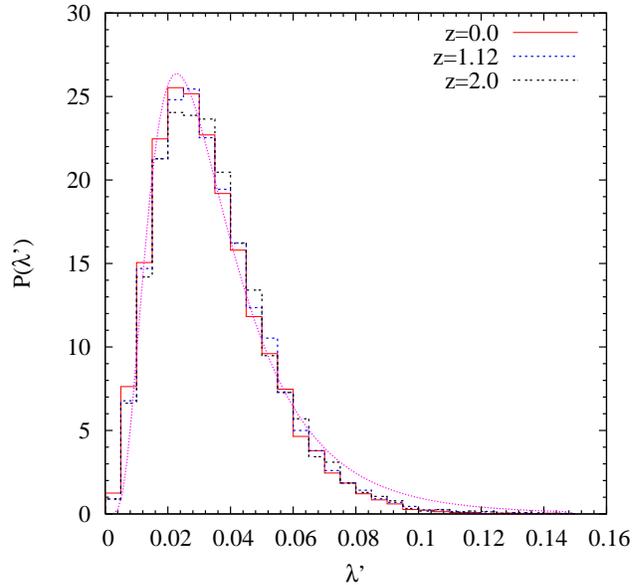}
  \caption{Distribution of halo spin parameter $\lambda'$ at redshifts
    $z=0,1,2$, compared to a log-normal distribution with
    $\lambda_0'=0.031$ and $\sigma=0.57$.}
  \label{spinhist}
\end{figure}

Finally, Fig.~\ref{spinhist} shows the distribution of spin parameters
at redshifts $z=0, 1.12$ and $2$. At all redshifts the distribution is
well fitted by a log-normal distribution,

\begin{equation}
P(\lambda')=\frac{1}{\lambda'\sqrt{2\pi}\sigma}\exp\left( - \frac{\ln^2(\lambda'/\lambda_0')}{2\sigma^2}\right),
\end{equation}
with $\sigma=0.57$ and $\lambda_0'=0.031$. This last quantity
$\lambda_0'$ shows a marginal mass dependence at high redshift as
shown in figure \ref{spin}.

\section{Discussion and Conclusions}
\label{sec:summ}

In this work we present a detailed analysis of a large set of N-body
simulations performed within a WMAP 5th year \LCDM cosmology (Komatsu
\etal 2009). We study the relation between structural properties of
dark matter haloes (concentration, spin and shape) with mass, and the
evolution of such scaling relations with redshift. We span the entire
mass range important for galaxy formation [$10^{10}:10^{15}
h^{-1}\,\rm M_{\odot}$] and a redshift range from $z=0$ to $z=2$.

We present results for ``relaxed'' haloes, defined according to the
criteria suggested by Macci\`o \etal (2007).  In our mass and redshift
range the \cM relation always follows a power law behavior.  We
confirmed that the redshift dependence of such relation is more
complex than a simple $(1+z)^{-1}$ scaling as proposed by Bullock
\etal 2001a, with both the normalization and the slope of the relation
changing with cosmic time.  We also found that for increasing
redshifts ($z \approx 2$) the power law behaviour seems to break in
agreement with recent studies (e.g.  Klypin \etal 2010).  Thanks to
our multiple box simulations we tested our results against resolution
effects and find them to be stable once a sufficient large number of
particles is used $N_{\rm vir}>500$.

Recently two other works have addressed the topic of the evolution of
the \cM relation, Zhao \etal (2009) and Klypin \etal (2010).  When
compared with the model proposed in Zhao \etal (2009) our results show
a very good agreement at the low mass end (possibly due to the fact
that both halo samples have more or less the same level of
resolution). For high masses we find a slightly higher difference but
never exceeding few percent.  The comparison with Klypin \etal is less
straightforward since they used a different method to compute
concentrations, based on the circular velocity of the halo instead of
directly fitting the density profile. Moreover they use all haloes in
their simulation volume without any distinction between relaxed and
unrelaxed.  Our results are in qualitative agreement with the model
proposed by Klypin \etal for the evolution of the \cM relation but
there are differences of the order of 8\%.  It is then interesting to
ask ourselves if these differences arise from the different method
used to compute $\cvir$.

For this purpose we applied to our B90 box (which has roughly the same
resolution as the Bolshoi simulation of Klypin \etal 2010) the method
proposed by Klypin to compute the concentration, based on the relation
between mass and maximum circular velocity. To be consistent with
Klypin \etal 2010 we considered all haloes in our sample, without
making any distinction between relaxed and unrelaxed.  We find that
the different methods to compute $c_{\rm vir}$ do not introduce any
systematic bias, being perfectly consistent.  Nevertheless we found
that at fixed mass our haloes have a lower circular velocity compared
to the Klypin \etal (2010) results. Taking into account the explicit
dependence of the concentration on $\rm{V_c}$ we are keen to conclude
that part of the difference in \cM relation could be due the slightly
different values of the cosmological parameters and primordial power
spectral index.

Another interesting result regards the effect of including unrelaxed
haloes.  Klypin \etal (2010) did not make any attempts of removing
unrelaxed haloes; this is in principle justified because their
concentrations are obtained from an integral quantity ($V_{\rm c}$)
which is less affected by the dynamical status of the halo with
respect to a differential one like $\rho$.  On the other hand by
applying Klypin's method to our B90 box we found that the $V_{\rm
  c}-M$ relation has a systematic shift ($\approx 6\%$) towards higher
values when it is computed for ``relaxed'' haloes instead than for all
haloes.  This results in a systematic bias, towards lower
concentrations, for the "all'' sample with respect to the relaxed one.

Finally Zhao \etal (2009) and Klypin \etal (2010) found that the
concentration may have a minimum value close to 3.5-4.5 at redshift
close to 4. Unfortunately our simulations do not have enough
resolution to get a statistical valid halo sample at such a high
redshift, so we cannot confirm such a finding, even if we do see some
evidence of a breaking of a simple power law behaviour for the \cM
relation at $z=2$.  Let us stress one more time that the fitting
formula we proposed in this work are valid only in the test redshift
range [0-2] and should not be extrapolated at higher redshifts.

In order to improve our understanding on the redshift evolution of the
\cM relation we look at the individual evolution with time of $\rs$
and $\Rvir$.  Both these length scales grow with decreasing redshift
until a maximum is reached, then they start to decrease towards
$z=0$. There is a clear analogy between the collapse of a linear
perturbation and the behaviour of $\rs$ and $\Rvir$. We found that we
can model the evolution of the inner part of the halo (within $\rs$)
as a decoupled spherical perturbation growing inside the central
region of the halo. The temporal offset between the ``turning points''
of the perturbations associated with $\rs$ and $\Rvir$ is able to
explain the observed redshift evolution of the \cM relation; which
strongly deviates from a simple $(1+z)^\alpha$ scaling of the $z=0$
relation. As a final remark we would like to stress that our results
refer to the dark matter distrubution in the absence of a collisional
component.  Although it has been shown that the inclusion of baryonic
physics may affect the properties of the dark matter, it also known
that the strength of this effect strongly depends on the implemented
baryonic physics (Duffy \etal 2010, Governato \etal 2010).  The dark
matter distribution in real haloes is still under debate and a
comparison of pure dark matter results with observations should be
preformed with extreme caution.

We then investigate the mass and redshift dependence of the axis ratio
($p$ and $s$) of dark matter haloes. Our results are in agreement with
previous studies and show that although on average haloes in our
simulations are preferentially triaxial at all masses and redshifts,
low mass haloes are more spherical than high mass ones and, at any
mass, central regions are more aspherical than outer ones. We find a
more complex evolution of the shape-mass relation with redshift with
respect to the model proposed by Allgood \etal (2006). We propose a
new fitting function (that deviates from a simple power law) that is
able to reproduce our data in the whole redshift and mass range.

Finally we studied the evolution of the spin parameter $\lambda'$.  At
redshift zero we confirm the results of Macci\'o \etal (2007; 2008),
who found the spin parameter to be mass independent. For increasing
redshift there is evidence of a correlation between halo mass and
spin: on average, more massive haloes have lower values of the spin
parameter. This is in agreement with recent findings at very high
redshift ($z=10$) by Knebe \& Power (2008).  As already speculated by
those authors, since disk sizes are to first order proportional to the
spin parameter (Mo \etal 1998), a lower spin parameter for high mass
haloes could make their central stellar and gaseous body more compact
and hence allowing more efficient star formation. In addition this
could affect the evolution of the size-mass relation of galaxy disks,
which is usually modeled by assuming $\lambda$ is independent of mass
and redshift (Mao, Mo \& White 1998; Somerville \etal 2008; Firmani \&
Avila-Reese 2009; Dutton \etal 2010).

However, in our simulations the differences in median spin parameters
at $z=2$ compared to $z=0$ are at most 15\%, and thus the consequences
to observables such as the size-mass relation of disk galaxies are
likely to be small.  Furthermore, baryonic effects such as supernova
driven outflows and inefficient cooling can modify spin parameters by
much larger amounts (factors of $\sim 2$) (Dutton \& van den Bosch
2009). Thus it is unlikely that the mass dependence on halo spin
that we observe in our simulations at $z=2$ will have an unambiguous
observational signature.

\section*{Acknowledgments}

The authors are in debt with Anatoly Klypin and Volker M\"{u}ller for
useful comments and discussions.  We acknowledge the LEA Astro-PF
collaboration and the ASTROSIM network of the European Science
Foundation (Science Meeting 2387) for the financial support of the
workshop ‘The Local Universe: From Dwarf Galaxies to Galaxy Clusters’
held in Jablonna near Warsaw in 2009 June/July, where part of this
work was done.  J.C.M. wants to give thanks to German Science
Foundation (grant MU 1020/6-4), to the Max-Planck-Institut f\"{u}r
Astronomie for its hospitality.  Numerical simulations were performed
on the PIA and on PanStarrs2 clusters of the Max-Planck-Institut f\"ur
Astronomie at the Rechenzentrum in Garching.



\section*{Appendix A: Detailed values for the parameters of the
  different fits presented in the main body of the paper.}
\label{sec:appendix}

We give in this appendix a complete set of tables with the value of all the
fitting parameters presented in the paper. All fits to the data are performed
for the functional form $\log(\psi) = a\log(\Mvir) + b$ where $\psi$ can be
either $\cvir$, $\lambda'$, $s$ or $p$ and where $\Mvir$ is in units of
$10^{10}\hMsun$. We stress on the validity of our results in the range of
redshifts between 0 and 2.  Nevertheless individual fits are valid in the
individual mass ranges $[10^{10}\hMsun, 10^{15}\hMsun]$ for z=0,
$[10^{10}\hMsun, \sim3.16^{14}\hMsun]$ for z=0.23, 0.38 and 0.56,
$[10^{10}\hMsun, \sim1.26^{14}\hMsun]$ for z=0.8, and 1.12 and
$[10^{10}\hMsun, \sim5.0^{13}\hMsun]$ for z=1.59, and 2.0.

We present in table \ref{tab:Table2}, just for completeness, the fitting of
the shape paramaters to $\log(\psi) = a\log(\Mvir) + b$ for all
masses and redshifts. In table \ref{tab:Table3} we present the fit to Eq.
\ref{eq:NewShape} to the shape parameters $s$,$p$ and $s_{0.3}$.

\begin{table}
\caption{Values of the fit parameters for the data presented in the
  paper. $\log(\cvir)$, $\log(\lambda')$ as computed from haloes with at least
  500 and 1000 particles inside the virial radius and $\log(s)$ and $\log(p)$
  as computed for haloes with more than 4000 particles inside the virial
  radius. All fits where done to the function $\log(\psi) = a\log(\Mvir) +
  b$. N haloes is the total number of relaxed haloes used at that redshift to
  compute the mean and median values.\label{tab:Table2}}
\begin{center}
\begin{tabular}{|c|c|c|c|c|c|}\hline \hline
Redshift & N haloes & a      & $\Delta$a & b     & $\Delta$b  \\ \hline \hline
\multicolumn{6}{c}{$\log(\cvir)$ (Nmin = 500)}\\
\\
0        & 23777   & -0.097 & 0.002     & 2.155 & 0.021  \\
0.23     & 22358   & -0.091 & 0.002     & 2.011 & 0.027  \\
0.38     & 20906   & -0.085 & 0.001     & 1.897 & 0.018  \\
0.56     & 19475   & -0.078 & 0.001     & 1.763 & 0.013  \\
0.8      & 17696   & -0.075 & 0.002     & 1.684 & 0.030  \\
1.12     & 14960   & -0.066 & 0.001     & 1.523 & 0.017  \\
1.59     & 11888   & -0.049 & 0.003     & 1.257 & 0.032  \\
2.0      & 9290    & -0.039 & 0.005     & 1.093 & 0.055  \\
\hline
\multicolumn{6}{c}{$\log(\lambda')$ (Nmin = 500)}\\
\\
0        & 23777   & -0.001 & 0.004     & -1.497 & 0.046 \\
0.23     & 22358   & -0.000 & 0.003     & -1.509 & 0.038 \\
0.38     & 20906   & -0.006 & 0.004     & -1.429 & 0.044 \\
0.56     & 19475   & -0.015 & 0.005     & -1.308 & 0.056 \\
0.8      & 17696   & -0.011 & 0.003     & -1.369 & 0.043 \\
1.12     & 14960   & -0.020 & 0.004     & -1.265 & 0.046 \\
1.59     & 11888   & -0.024 & 0.004     & -1.220 & 0.048 \\
2.0      & 9290    & -0.030 & 0.005     & -1.150 & 0.064 \\
\hline
\multicolumn{6}{c}{$\log(\lambda')$ (Nmin = 1000)}\\
\\
0        & 12427   & -0.002 & 0.005     & -1.485 & 0.063 \\
0.23     & 11563   & -0.004 & 0.004     & -1.452 & 0.056 \\
0.38     & 10753   & -0.010 & 0.005     & -1.373 & 0.066 \\
0.56     & 10049   & -0.012 & 0.005     & -1.337 & 0.062 \\
0.8      & 8942    & -0.013 & 0.005     & -1.344 & 0.065 \\
1.12     & 7482    & -0.016 & 0.005     & -1.292 & 0.058 \\
1.59     & 5851    & -0.030 & 0.005     & -1.146 & 0.067 \\
2.0      & 4404    & -0.035 & 0.011     & -1.091 & 0.137 \\
\hline
\multicolumn{6}{c}{$\log(s)$ (Nmin = 4000)}\\
\\
0        & 3102   & -0.039  &  0.006    & 0.386  & 0.079 \\
0.23     & 2785   & -0.033  &  0.003    & 0.294  & 0.044 \\
0.38     & 2523   & -0.036  &  0.004    & 0.322  & 0.059 \\
0.56     & 2281   & -0.035  &  0.003    & 0.303  & 0.034 \\
0.8      & 1972   & -0.038  &  0.004    & 0.333  & 0.055 \\
1.12     & 1588   & -0.033  &  0.005    & 0.260  & 0.063 \\
1.59     & 1211   & -0.025  &  0.005    & 0.153  & 0.059 \\
2.0      & 879    & -0.024  &  0.010    & 0.133  & 0.123 \\
\hline
\multicolumn{6}{c}{$\log(p)$ (Nmin = 4000)}\\
\\
0        & 3102   & -0.051  & 0.007     & 0.443  & 0.096 \\
0.23     & 2785   & -0.044  & 0.005     & 0.344  & 0.062 \\
0.38     & 2523   & -0.043  & 0.005     & 0.318  & 0.065 \\
0.56     & 2281   & -0.044  & 0.004     & 0.314  & 0.051 \\
0.8      & 1972   & -0.044  & 0.006     & 0.298  & 0.071 \\
1.12     & 1588   & -0.039  & 0.006     & 0.224  & 0.076 \\
1.59     & 1211   & -0.034  & 0.005     & 0.145  & 0.059 \\
2.0      & 879    & -0.032  & 0.012     & 0.112  & 0.147 \\
\hline
\hline
\end{tabular}
\end{center}
\end{table}

\begin{table}
\caption{Values of the fit parameters for the data presented in the
  paper $s$, $p$ and $s_{0.3}$ as computed for haloes with more than
  1000 and 4000 particles inside the virial radius respectivelly. All
  fits where done to the function $\psi = \alpha[\log(\Mvir)]^4 +
  \beta$. N haloes is the total number of relaxed haloes used at that
  redshift to compute the mean and median values.\label{tab:Table3}}
\begin{center}
\begin{tabular}{|c|c|c|c|c|c|}\hline \hline
Redshift & N haloes & $\alpha (\times10^{-6})$     & $\Delta \alpha (\times10^{-7})$ & $\beta$   & $\Delta \beta$  \\ \hline \hline
\multicolumn{6}{c}{$s$ (Nmin = 1000)}\\
\\
0.0     & 12426 & -6.566 & 2.760 & 0.815 & 0.008 \\
0.23    & 11563 & -7.622 & 2.818 & 0.815 & 0.008 \\
0.38    & 10753 & -7.009 & 1.185 & 0.783 & 0.003 \\
0.56    & 10049 & -7.215 & 1.484 & 0.771 & 0.004 \\
0.8     & 8942  & -7.173 & 3.320 & 0.751 & 0.008 \\
1.12    & 7482  & -6.388 & 3.945 & 0.714 & 0.010 \\
1.59    & 5851  & -5.932 & 3.760 & 0.677 & 0.008 \\
2.0     & 4404  & -5.512 & 3.936 & 0.655 & 0.009 \\
\hline
\multicolumn{6}{c}{$p$ (Nmin = 1000)}\\
\\
0.0     & 12426 & -6.345 & 2.510 & 0.955 & 0.007 \\
0.23    & 11563 & -7.396 & 7.872 & 0.955 & 0.023 \\
0.38    & 10753 & -6.452 & 2.024 & 0.923 & 0.005 \\
0.56    & 10049 & -6.802 & 1.755 & 0.916 & 0.005 \\
0.8     & 8942  & -7.539 & 2.984 & 0.916 & 0.007 \\
1.12    & 7482  & -5.992 & 5.600 & 0.867 & 0.014 \\
1.59    & 5851  & -6.299 & 4.812 & 0.845 & 0.011 \\
2.0     & 4404  & -5.118 & 7.149 & 0.808 & 0.016 \\
\hline 
\multicolumn{6}{c}{$s_{0.3}$ (Nmin = 4000)}\\
\\
0       & 3102 & -6.464 & 2.709 & 0.760 & 0.008 \\
0.23    & 2785 & -7.484 & 3.404 & 0.760 & 0.010 \\
0.38    & 2523 & -6.776 & 2.803 & 0.722 & 0.008 \\
0.56    & 2281 & -6.870 & 3.522 & 0.703 & 0.010 \\
0.8     & 1972 & -6.444 & 3.397 & 0.682 & 0.009 \\
1.12    & 1588 & -5.487 & 5.210 & 0.632 & 0.014 \\
1.59    & 1211 & -6.441 & 7.297 & 0.632 & 0.020 \\
2.0     & 879  & -4.050 & 0.138 & 0.562 & 0.035 \\

\hline
\hline
\end{tabular}
\end{center}
\end{table}

\bsp

\label{lastpage}

\end{document}